\documentclass{elsart}

\def  \betabf  {{\mbox{\boldmath$\beta$}}}
\def  \thetabf  {{\mbox{\boldmath$\theta$}}}

\usepackage{natbib}
\usepackage{lscape}
\usepackage{rotating}

\journal{Accident Analysis and Prevention}\date{}

\begin{document}

\begin{frontmatter}


\title{Markov switching negative binomial models:
an application to vehicle accident frequencies}

\author{Nataliya V. Malyshkina\corauthref{correspond_author}},
\corauth[correspond_author]{Corresponding author.}
\ead{nmalyshk@purdue.edu}
\author{Fred L. Mannering},
\ead{flm@ecn.purdue.edu}
\author{Andrew P. Tarko}
\ead{tarko@ecn.purdue.edu}
\address{School of Civil Engineering, 550 Stadium Mall Drive,
Purdue University, West Lafayette, IN 47907, United States}

\begin{abstract}
In this paper, two-state Markov switching models are proposed to
study accident frequencies. These models assume that there are
two unobserved states of roadway safety, and that roadway entities
(roadway segments) can switch between these states over time.
The states are distinct, in the sense that in the different
states accident frequencies are generated by separate counting
processes (by separate Poisson or negative binomial processes).
To demonstrate the applicability of the approach presented
herein, two-state Markov switching negative binomial models
are estimated using five-year accident frequencies on selected
Indiana interstate highway segments. Bayesian inference methods
and Markov Chain Monte Carlo (MCMC) simulations are used for
model estimation. The estimated Markov switching models result
in a superior statistical fit relative to the standard
(single-state) negative binomial model. It is found that the more
frequent state is safer and it is correlated with better weather
conditions. The less frequent state is found to be less safe and
to be correlated with adverse weather conditions.
\end{abstract}


\end{frontmatter}

\section{Introduction}
\label{INTRO}

Vehicle accidents place an
incredible social and economic burden on society. As a result,
considerable research has been conducted on understanding and
predicting accident frequencies (the number of accidents
occurring on roadway segments over a given time period). Because
accident frequencies are non-negative integers, count data models are
a reasonable statistical modeling approach \citep[][]{WKM_03}. Simple
modeling approaches include Poisson models and negative binomial (NB)
models \citep[][]{HALW_95,SMB_95,PM_96,ML_03}. These models assume
a single process for accident data generation (a Poisson process
or a negative binomial process) and involve a nonlinear regression of
the observed accident frequencies on various roadway-segment
characteristics (such as roadway geometric and environmental
factors). Because a preponderance of zero-accident observations is
often observed in empirical data, \citet[][]{M_94}, \citet[][]{SMM_97}
and others have applied zero-inflated Poisson (ZIP) and
zero-inflated negative binomial (ZINB) models for predicting accident
frequencies. Zero-inflated models assume a two-state process for
accident data generation -- one state is assumed to be safe with zero
accidents (over the duration of time being considered) and the other
state is assumed to be unsafe with a possibility of nonzero accident
frequencies. In zero-inflated models, individual roadway segments are
assumed to be always in the safe or unsafe state. While
the application of zero-inflated models often provides a better
statistical fit of observed accident frequency data, the
applicability of these models has been questioned by
\citet[][2007]{LWI_05}. In particular, \citet[][2007]{LWI_05} argue
that it is unreasonable to expect some roadway segments to be
always perfectly safe. In addition, they argue that zero-inflated
models do not account for a likely possibility for roadway segments
to change in time from one state to another.

\begin{figure}[t]
\vspace{4.75truecm}
\includegraphics{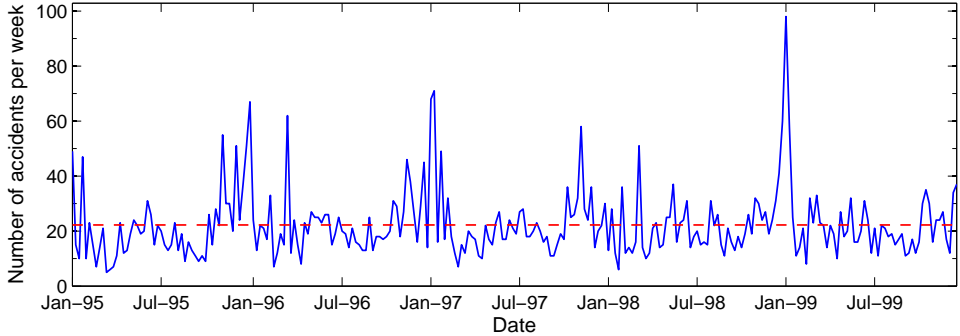}
\caption{Weekly accident frequencies on the sample of Indiana
interstate segments from 1995 to 1999 (the horizontal dashed
line shows the average value).
\label{FIGURE_1}
}
\end{figure}

In this paper, two-state Markov switching count data models are
explored as a method for studying accident frequencies. These models
assume Markov switching (over time) between two unobserved states of
roadway safety.\footnote{In fact, there may be more than two states
but, for illustration purposes, the two-state case will be considered
in this paper. Extending the approach to the possibility of additional
states would significantly complicate the model structure. However,
once this extension were done, additional states could be empirically
tested.}
There are a number of reasons to expect the existence of multiple
states. First, the safety of roadway segments is likely to vary under
different environmental conditions, driver reactions and other factors
that may not necessarily be available to the analyst. For an
illustration, consider Figure~\ref{FIGURE_1}, which shows weekly
time series of the number of accidents on selected Indiana interstate
segments during the 1995-1999 time interval. The figure shows that the
number of accidents per week fluctuates widely over time. Thus, under
different conditions, roads can become considerably more or less safe.
As a result, it is reasonable to assume that there exist two or more
states of roadway safety. These states can help account for the existence
of numerous unidentified and/or unobserved factors (unobserved
heterogeneity) that influence roadway safety. Markov switching models are
designed to account for unobserved states in a convenient and feasible
way. In fact, these models have been successfully applied in other
scientific fields. For example, two-state Markov switching
autoregressive models have been used in economics, where the two
states are usually identified as economic recession and expansion
\citep[][]{H_89,MT_94,T_02}.

Another reason to expect the existence of multiple states is the
empirical success of zero-inflated models, which are predicated on
the existence of two-state process -- a safe and an unsafe state
\citep[see][]{SMM_97,CM_01,LM_02}. Markov switching can be viewed
as an extension of previous work on zero-inflated models, in that
it relaxes the assumption that a safe state exists (which has been
brought up as a concern, see \citet[][2007]{LWI_05}) and instead
considers two significantly different unsafe states. In addition,
in contrast to zero-inflated models, Markov switching explicitly
considers the possibility that roadway segments can change their
state over time.

\section{Model specification}
\label{MOD_SPECIF}

Markov switching models are parametric and can be fully
specified by a likelihood function $f({\bf Y}|{\bf\Theta},{\cal M})$,
which is the conditional probability distribution
of the vector of all observations ${\bf Y}$, given the vector of all
parameters ${\bf\Theta}$ of model ${\cal M}$. In our study, we
observe the number of accidents $A_{t,n}$ that occur on the $n^{\rm th}$
roadway segment during time period $t$. Thus ${\bf Y}=\{A_{t,n}\}$
includes all accidents observed on all roadway segments over all
time periods ($n=1,2,\ldots,N_t$ and $t=1,2,\ldots,T$). Model
${\cal M}=\{M,{\bf X}_{t,n}\}$ includes the model's name $M$ (for
example, $M=\mbox{``negative binomial''}$) and the vector
${\bf X}_{t,n}$ of all roadway segment characteristic variables
(segment length, curve characteristics, grades, pavement properties,
and so on).

To define the likelihood function, we first
introduce an unobserved (latent) state variable $s_t$, which
determines the state of all roadway segments during time period $t$.
At each $t$, the state variable $s_t$ can assume only two values:
$s_t=0$ corresponds to one state and $s_t=1$ corresponds to the
other state. The state variable $s_t$ is assumed to follow a
stationary two-state Markov chain process in
time,\footnote{
\label{MARKOV} Markov property means that the probability
distribution of $s_{t+1}$ depends only on the value $s_{t}$ at time $t$,
but not on the previous history $s_{t-1},s_{t-2},\ldots$
\citep[][]{B_69}. Stationarity of $\{s_t\}$ is in the statistical sense.
Below we will relax the assumption of stationarity and discuss a case of
time-dependent transition probabilities.}
which can be specified by time-independent transition probabilities as
\begin{eqnarray}
P(s_{t+1}=1|s_t=0)=p_{0\rightarrow1},
\quad P(s_{t+1}=0|s_t=1)=p_{1\rightarrow0}.
\label{EQ_P}
\end{eqnarray}
Here, for example, $P(s_{t+1}=1|s_t=0)$ is the conditional probability
of $s_{t+1}=1$ at time $t+1$, given that $s_t=0$ at time $t$. Note that
$P(s_{t+1}=0|s_t=0)=1-p_{0\rightarrow1}$ and
$P(s_{t+1}=1|s_t=1)=1-p_{1\rightarrow0}$. Transition probabilities
$p_{0\rightarrow1}$ and $p_{1\rightarrow0}$ are unknown parameters to
be estimated from accident data. The stationary unconditional
probabilities $\bar p_{0}$ and $\bar p_{1}$ of states $s_t=0$ and
$s_t=1$ are\footnote{
These can be found from stationarity conditions
$\bar p_0=(1-p_{0\rightarrow1})\bar p_0+p_{1\rightarrow0}\bar p_1$,
$\bar p_1=p_{0\rightarrow1}\bar p_0+(1-p_{1\rightarrow0})\bar p_1$ and
$\bar p_0+\bar p_1=1$ \citep[][]{B_69}.}
\begin{eqnarray}
\begin{array}{lcl}
\bar p_0 & = & p_{1\rightarrow0}/(p_{0\rightarrow1}+p_{1\rightarrow0})
\quad\quad\mbox{for state}\quad\quad s_t=0,\\
\bar p_1 & = & p_{0\rightarrow1}/(p_{0\rightarrow1}+p_{1\rightarrow0})
\quad\quad\mbox{for state}\quad\quad s_t=1.
\label{EQ_P_BAR}
\end{array}
\end{eqnarray}
Without loss of generality, we assume that (on average) state $s_t=0$
occurs more or equally frequently than state $s_t=1$.
Therefore,
$\bar p_{0}\geq\bar p_{1}$, and from Eqs.~(\ref{EQ_P_BAR}) we obtain
restriction\footnote{Restriction~(\ref{EQ_P_RESTRICT}) allows to avoid
the problem of switching of state labels, \mbox{$0\leftrightarrow1$}.
This problem would otherwise arise because of the symmetry of the
likelihood function~(\ref{EQ_L_1})--(\ref{EQ_L}) under the label
switching.}
\begin{eqnarray}
p_{0\rightarrow1}\leq p_{1\rightarrow0}.
\label{EQ_P_RESTRICT}
\end{eqnarray}
We refer to states $s_t=0$ and $s_t=1$ as ``more frequent''
and ``less frequent'' states respectively.

Next, consider a two-state Markov switching negative binomial (MSNB)
model that assumes negative binomial data-generating processes in each
of the two states. With this, the probability of $A_{t,n}$ accidents
occurring on roadway segment $n$ during time period $t$ is
\begin{eqnarray}
P_{t,n}^{(A)} & = & \displaystyle{\frac{\Gamma(A_{t,n}+1/\alpha_t)}
{\Gamma(1/\alpha_t)A_{t,n}!}
\left(\frac{1}{1+\alpha_t\lambda_{t,n}}\right)^{1/\alpha_t}
\left(\frac{\alpha_t\lambda_{t,n}}{1+\alpha_t\lambda_{t,n}}\right)^{A_{t,n}}},
\label{EQ_L_1}\\
\lambda_{t,n} & = & \left\{
    \begin{array}{lll}
    \exp(\betabf_{(0)}'{\bf X}_{t,n}) & \quad\mbox{if}\quad & s_t=0\\
    \exp(\betabf_{(1)}'{\bf X}_{t,n}) & \quad\mbox{if}\quad & s_t=1
    \end{array}
\right.,
\label{EQ_LAMBDA}\\
\alpha_t & = & \left\{
    \begin{array}{lll}
    \alpha_{(0)} & \quad\mbox{if}\quad & s_t=0\\
    \alpha_{(1)} & \quad\mbox{if}\quad & s_t=1
    \end{array}
\right.,
\nonumber\\
t &=& 1,2,\ldots,T,\quad n=1,2,\ldots,N_t. \vphantom{\int}
\nonumber
\end{eqnarray}
Here, Eq.~(\ref{EQ_L_1}) is the standard negative binomial probability
mass function \citep[][]{WKM_03}, $\Gamma(\;)$ is the gamma function,
prime means transpose (so $\betabf_{(0)}'$ is the transpose of
$\betabf_{(0)}$), $N_t$ is the number of roadway segments observed
during time period $t$, and $T$ is the total number of time periods.
Parameter vectors $\betabf_{(0)}$ and
$\betabf_{(1)}$, and over-dispersion parameters $\alpha_{(0)}\ge0$ and
$\alpha_{(1)}\ge0$ are the unknown estimable parameters of negative
binomial models in the two states, $s_t=0$ and $s_t=1$
respectively.\footnote{To ensure that $\alpha_{(0)}$ and
$\alpha_{(1)}$ are non-negative, their logarithms are used in
estimation.}
We set the first component of ${\bf X}_{t,n}$ to unity, and,
therefore, the first components of $\betabf_{(0)}$ and $\betabf_{(1)}$
are the intercepts in the two states.

If accident events are assumed to be independent, the likelihood
function is
\begin{eqnarray}
f({\bf Y}|{\bf\Theta},{\cal M})=\prod\limits_{t=1}^T\prod\limits_{n=1}^{N_t}P_{t,n}^{(A)}.
\label{EQ_L}
\end{eqnarray}
Here, because the state variables $s_t$ are unobservable, the vector
of all estimable parameters ${\bf\Theta}$ must include all states, in
addition to all model parameters ($\beta$-s, $\alpha$-s) and
transition probabilities. Thus,
\begin{eqnarray}
{\bf\Theta}=[\betabf_{(0)}',\alpha_{(0)},\betabf_{(1)}',\alpha_{(1)},p_{0\rightarrow1},
p_{1\rightarrow0},{\bf S}']',\quad\quad{\bf S}'=[s_1,\ldots,s_T].
\label{EQ_THETA}
\end{eqnarray}
Vector ${\bf S}$ has length $T$ and contains all state values. Eqs.~(\ref{EQ_P})-(\ref{EQ_THETA}) define the two-state Markov
switching negative binomial (MSNB) models considered in this study.

\section{Model estimation methods}
\label{MOD_ESTIM}

Statistical estimation of Markov switching models is complicated by
unobservability of the state variables
$s_t$.\footnote{\label{FN_S}Below we will have 260 time periods
($T=260$). In this case, there are $2^{260}$ possible
combinations for value of vector ${\bf S}=[s_1,\ldots,s_T]'$.}
As a result, the traditional maximum likelihood estimation (MLE)
procedure is of very limited use for Markov switching models.
Instead, a Bayesian inference approach is used.
Given a model ${\cal M}$ with likelihood function
$f({\bf Y}|{\bf\Theta},{\cal M})$, the Bayes formula is
\begin{eqnarray}
f({\bf\Theta}|{\bf Y},{\cal M})=
\frac{f({\bf Y},{\bf\Theta}|{\cal M})}{f({\bf Y}|{\cal M})}=
\frac{f({\bf Y}|{\bf\Theta},{\cal M})\pi({\bf\Theta}|{\cal M})}
{\int f({\bf Y},{\bf\Theta}|{\cal M})\,d{\bf\Theta}}.
\label{EQ_POSTERIOR}
\end{eqnarray}
Here $f({\bf\Theta}|{\bf Y},{\cal M})$ is the posterior probability
distribution of model parameters ${\bf\Theta}$ conditional on the
observed data ${\bf Y}$ and model ${\cal M}$.
Function $f({\bf Y},{\bf\Theta}|{\cal M})$ is the
joint probability distribution of ${\bf Y}$ and ${\bf\Theta}$ given
model ${\cal M}$. Function $f({\bf Y}|{\cal M})$ is the marginal
likelihood function -- the probability distribution of data
${\bf Y}$ given model ${\cal M}$. Function $\pi({\bf\Theta}|{\cal M})$
is the prior probability distribution of parameters that reflects
prior knowledge about ${\bf\Theta}$. The intuition behind
Eq.~(\ref{EQ_POSTERIOR}) is straightforward: given model ${\cal M}$,
the posterior distribution accounts for both the observations ${\bf Y}$
and our prior knowledge of ${\bf\Theta}$. We use the harmonic mean
formula to calculate the marginal likelihood $f({\bf Y}|{\cal M})$
of data ${\bf Y}$ \citep[see][]{KR_95} as,
\begin{eqnarray}
f({\bf Y}|{\cal M})^{-1}=\int\frac{f({\bf\Theta}|{\bf Y},{\cal M})}{f({\bf Y}|{\bf\Theta},{\cal M})}
\,d{\bf\Theta}=E\left[\left.f({\bf Y}|{\bf\Theta},{\cal M})^{-1}\right|{\bf Y}\right],
\label{EQ_L_MARGINAL}
\end{eqnarray}
where $E(\ldots|{\bf Y})$ is the posterior expectation (which
is calculated by using the posterior distribution).

In our study (and in most practical studies), the direct application
of Eq.~(\ref{EQ_POSTERIOR}) is not feasible because the parameter vector
${\bf\Theta}$ contains too many components, making integration over
${\bf\Theta}$ in Eq.~(\ref{EQ_POSTERIOR}) extremely difficult. However,
the posterior distribution $f({\bf\Theta}|{\bf Y},{\cal M})$ in
Eq.~(\ref{EQ_POSTERIOR}) is known up to its normalization constant,
$f({\bf\Theta}|{\bf Y},{\cal M})\propto
f({\bf Y}|{\bf\Theta},{\cal M})\pi({\bf\Theta}|{\cal M})$. As a result,
we use Markov Chain Monte Carlo (MCMC) simulations, which provide a
convenient and practical computational methodology for sampling
from a probability distribution known up to a constant (the posterior
distribution in our case). Given a large enough posterior sample of
parameter vector ${\bf\Theta}$, any posterior expectation and variance
can be found and Bayesian inference can be readily applied. In the
Appendix we describe our choice of prior distribution
$\pi({\bf\Theta}|{\cal M})$ and the MCMC simulations. The
prior distribution is chosen to be wide and essentially
noninformative.
For the MCMC simulations in this paper, special numerical code
was written in the MATLAB programming language and tested on
artificial accident data sets. The test procedure included a generation
of artificial data with a known model. Then these data were used
to estimate the underlying model by means of our simulation code. With
this procedure we found that the MSNB models, used to generate the
artificial data, were reproduced successfully with our estimation
code.

For comparison of different models we use the following Bayesian
approach. Let there be two models ${\cal M}_1$ and ${\cal M}_2$ with
parameter vectors ${\bf\Theta_1}$ and ${\bf\Theta_2}$
respectively. Assuming that we have equal preferences of these models,
their prior probabilities are $\pi({\cal M}_1)=\pi({\cal
M}_2)=1/2$. In this case, the ratio of the models' posterior
probabilities, $P({\cal M}_1|{\bf Y})$ and $P({\cal M}_2|{\bf Y})$, is
equal to the Bayes factor. The later is defined as the ratio of the
models' marginal likelihoods \citep[][]{KR_95}. Thus, we have
\begin{eqnarray}
\frac{P({\cal M}_2|{\bf Y})}{P({\cal M}_1|{\bf Y})}=\frac{f({\cal M}_2,{\bf Y})/f(\bf Y)}
{f({\cal M}_1,{\bf Y})/f(\bf Y)}=
\frac{f({\bf Y}|{\cal M}_2)\pi({\cal M}_2)}{f({\bf Y}|{\cal M}_1)\pi({\cal M}_1)}=
\frac{f({\bf Y}|{\cal M}_2)}{f({\bf Y}|{\cal M}_1)},
\label{EQ_BAYES_FACTOR}
\end{eqnarray}
where $f({\cal M}_1,{\bf Y})$ and $f({\cal M}_2,{\bf Y})$ are the
joint distributions of the models and the data, $f({\bf Y})$ is the
unconditional distribution of the data, and the marginal likelihoods
$f({\bf Y}|{\cal M}_1)$ and $f({\bf Y}|{\cal M}_2)$ are given by
Eq.~(\ref{EQ_L_MARGINAL}). If the ratio in Eq.~(\ref{EQ_BAYES_FACTOR})
is larger than one, then model ${\cal M}_2$ is favored, if
the ratio is less than one, then model ${\cal M}_1$ is favored. An
advantage of the use of Bayes factors is that it has an inherent
penalty for including too many parameters in the model and
guards against overfitting.

\section{Model estimation results}
\label{RESULTS}

Data are used from 5769 accidents that were
observed on 335 interstate highway segments in
Indiana in 1995-1999. We use weekly time periods,
$t=1,2,3,\ldots,T=260$ in total.\footnote{A week is from Sunday to
Saturday, there are 260 full weeks in the 1995-1999 time interval.
We also considered daily time periods and obtained qualitatively
similar results (not reported here).}
Thus, in the present study the state ($s_t$) is
the same for all roadway segments and can change every week.
Four types of accident frequency models are estimated:
\begin{itemize}
\item
First, we estimate a standard (single-state) negative binomial (NB)
model without Markov switching by maximum likelihood estimation
(MLE). We refer to this model as ``NB-by-MLE''.
\item
Second, we estimate the same standard negative binomial model by the
Bayesian inference approach and the MCMC simulations. We refer to this
model as ``NB-by-MCMC''.
As one expects, for our choice of a
non-informative prior distribution, the estimated NB-by-MCMC
model turned out to be very similar to the NB-by-MLE model.
\item
Third, we estimate a restricted two-state Markov switching negative
binomial (MSNB) model. In this restricted switching model only the
intercept in the model parameters vector $\betabf$ and the
over-dispersion parameter $\alpha$ are allowed to switch between the
two states of roadway safety. In other words, in Eq.~(\ref{EQ_LAMBDA})
only the first components of vectors $\betabf_{(0)}$ and $\betabf_{(1)}$
may differ, while the remaining components are restricted to be the
same. In this case, the two states can have different average accident
rates, given by Eq.~(\ref{EQ_LAMBDA}), but the rates have the same
dependence on the explanatory variables. We refer to this model as
``restricted MSNB''; it is estimated by the Bayesian-MCMC methods.
\item
Fourth, we estimate a full two-state Markov switching negative
binomial (MSNB) model. In this model all estimable model parameters
($\beta$-s and $\alpha$) are allowed to switch between the two states
of roadway safety. To obtain the final full MSNB model reported here,
we consecutively construct and use $60\%$, $85\%$ and $95\%$ Bayesian
credible intervals for evaluation of the statistical significance of
each $\beta$-parameter. As a result, in the final model some
components of $\betabf_{(0)}$ and $\betabf_{(1)}$ are restricted to
zero or restricted to be the same in the two states.\footnote{A
$\beta$-parameter is restricted to zero if it is statistically
insignificant. A $\beta$-parameter is restricted to be the same in
the two states if the difference of its values in the two states
is statistically insignificant. A $(1-a)$ credible interval is chosen
in such way that the posterior probabilities of being below and above
it are both equal to $a/2$ (we use significance levels
$a=40\%,15\%,5\%$).}
We do not impose any restrictions on over-dispersion parameters
($\alpha$-s). We refer to the final full MSNB model as
``full MSNB''; it is estimated by the Bayesian-MCMC methods.
\end{itemize}
Note that the two states, and thus the MSNB models, do not have
to exist. For example, they will not exist if all estimated model
parameters turn out to be statistically the same in the two states,
$\betabf_{(0)}=\betabf_{(1)}$, (which suggests the two states are
identical and the MSNB models reduce to the standard NB model).
Also, the two states will not exist if all estimated state
variables $s_t$ turn out to be close to zero, resulting in
$p_{0\rightarrow1}\ll p_{1\rightarrow0}$ (compare to
Eq.~(\ref{EQ_P_RESTRICT})), then the less frequent state
$s_t=1$ is not realized and the process stays in state $s_t=0$.

The model estimation results for accident frequencies
are given in Table~\ref{T_1}.
Posterior (or MLE) estimates of all continuous model parameters
($\beta$-s, $\alpha$, $p_{0\rightarrow1}$ and $p_{1\rightarrow0}$)
are given together with the corresponding $95\%$ confidence intervals
for MLE models and $95\%$ credible intervals for Bayesian-MCMC models
(refer to the superscript and subscript numbers adjacent to parameter
posterior/MLE estimates in Table~\ref{T_1}).\footnote{Note
that MLE estimation assumes asymptotic normality of the estimates,
resulting in confidence intervals being symmetric around the means
(a $95\%$ confidence interval is $\pm1.96$ standard deviations around
the mean). In contrast, Bayesian estimation does not require this
assumption, and posterior distributions of parameters and Bayesian
credible intervals are usually non-symmetric.}
Table~\ref{T_2} gives summary statistics of all roadway segment
characteristic variables ${\bf X}_{t,n}$ (except the intercept).

\begin{figure}[t]
\vspace{9.5truecm}
\includegraphics{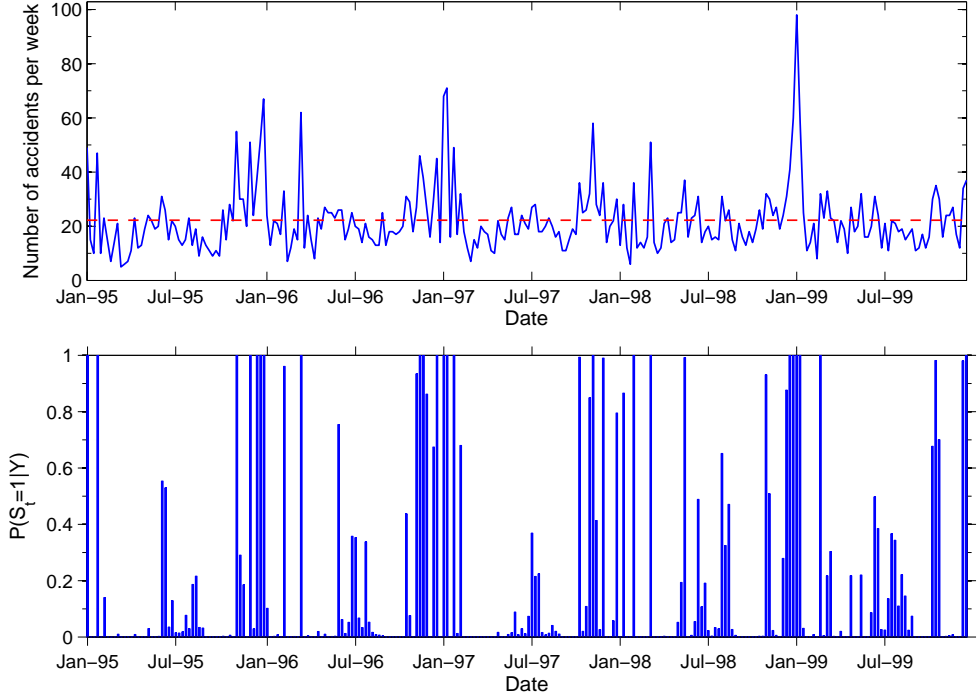}
\caption{The top plot is the same as Figure~\ref{FIGURE_1}.
The bottom plot shows weekly posterior probabilities
$P(s_t=1|{\bf Y})$ for the full MSNB model.
\label{FIGURE_2}
}
\end{figure}

To visually see how the model tracks the data, consider
Figure~\ref{FIGURE_2}. The top plot in Figure~\ref{FIGURE_2}
shows the weekly accident frequencies in our data as given in
Figure~\ref{FIGURE_1}. The bottom plot in Figure~\ref{FIGURE_2}
shows corresponding weekly posterior probabilities $P(s_t=1|{\bf Y})$
of the less frequent state $s_t=1$ for the full MSNB model. These
probabilities are equal to the posterior expectations of $s_t$,
$P(s_t=1|{\bf Y})=1\times P(s_t=1|{\bf Y})+0\times P(s_t=0|{\bf Y})=E(s_t|{\bf Y})$.
Weekly values of $P(s_t=1|{\bf Y})$ for the
restricted MSNB model are very similar to those given on the
top plot in Figure~\ref{FIGURE_2}, and, as a result, are not shown
on a separate plot. Indeed, the time-correlation\footnote{Here and
below we calculate weighted correlation coefficients. For variable
$P(s_t=1|{\bf Y})\equiv E(s_t|{\bf Y})$ we use weights $w_t$ inversely
proportional to the posterior standard deviations of $s_t$. That is
$w_t\propto \min\left\{1/{\rm std}(s_t|{\bf Y}),
{\rm median}[1/{\rm std}(s_t|{\bf Y})]\right\}$.
}
between $P(s_t=1|{\bf Y})$ for the two MSNB models is about
$99.5\%$.

Turning to the estimation results, the findings show that two
states exist and Markov switching models are strongly favored by
the empirical data. In particular, in the restricted MSNB model
we over $99.9\%$ confident that the difference in values of
$\beta$-intercept in the two states is
non-zero.\footnote{\label{INTERCEPTS}The difference of the intercept
values is statistically non-zero despite the fact that the $95\%$
credible intervals for these values overlap (see the ``Intercept''
line and the ``Restricted MSNB'' columns in Table~\ref{T_1}).
The reason is that the posterior draws of the intercepts are
correlated. The statistical test of whether the intercept values
differ, must be based on evaluation of their difference.}
To compare the Markov switching models (restricted and full MSNB) with
the corresponding standard non-switching model (NB), we calculate and
use Bayes factors given by Eq.~(\ref{EQ_BAYES_FACTOR}). We use
Eq.~(\ref{EQ_L_MARGINAL}) and bootstrap simulations\footnote{During
bootstrap simulations we repeatedly draw, with replacement, posterior
values of ${\bf\Theta}$ to calculate the posterior expectation in
Eq.~(\ref{EQ_L_MARGINAL}). In each of $10^5$ bootstrap draws that we
make, the number of ${\bf\Theta}$ values drawn is 1/100 of the total
number of all posterior ${\bf\Theta}$ values available from MCMC
simulations.}
for calculation of the values and the $95\%$ confidence intervals of
the logarithms of the marginal likelihoods given in Table~\ref{T_1}.
The log-marginal-likelihoods are $-16108.6$,
$-15850.2$ and $-15809.4$ for the NB, restricted MSNB and
full MSNB models respectively. Therefore, the restricted and
full MSNB models provide considerable ($258.4$ and $299.2$)
improvements of the log-marginal-likelihoods of the data as
compared to the corresponding standard non-switching NB model.
Thus, given the accident data, the posterior probabilities of
the restricted and full MSNB models are larger than the
probability of the standard NB model by $e^{258.4}$ and
$e^{299.2}$ respectively.

We can also use a classical statistics approach for model comparison,
based on the maximum likelihood estimation (MLE). Referring
to Table~\ref{T_1}, the MLE gives the maximum
log-likelihood value $-16081.2$ for the standard NB model. The
maximum log-likelihood values observed during our MCMC simulations
for the restricted and full MSNB models are $-15786.6$ and
$-15744.8$ respectively. An imaginary MLE, at its convergence, would
give MSNB log-likelihood values that were even larger than these
observed values. Therefore, the MSNB models provide very large
(at least $294.6$ and $336.4$) improvements in the maximum
log-likelihood value over the standard NB model. These improvements
come with only modest increases in the number of free continuous
model parameters ($\beta$-s and $\alpha$-s) that enter the
likelihood function. Both the Akaike Information Criterion (AIC)
and the Bayesian Information Criterion (BIC)\footnote{Minimization
of $AIC=2K-2LL$ and $BIC=K\ln(N)-2LL$ ensures an optimal choice
of explanatory variables in a model and avoids overfitting
\citep[][]{T_02,WKM_03}. Here $K$ is the number of free continuous
model parameters that enter the likelihood function, $N$ is the
number of observations and $LL$ is the log-likelihood. When
$N\geq8$, BIC favors fewer free parameters than AIC does.}
strongly favor the MSNB models over the NB model.

\begin{landscape}
\begin{table}[p]
\caption{Estimation results for negative binomial models of
accident frequency (the superscript and subscript numbers to the right of}
{individual parameter posterior/MLE estimates are $95\%$ confidence/credible
intervals -- see text for further explanation)}
\label{T_1}
\begin{scriptsize}

\tabcolsep=0.45em
\renewcommand{\arraystretch}{1.45}
\begin{tabular}{|l|c|c|c|c|c|c|}
\hline
${}_{{{}\atop{\displaystyle\mbox{\bf Variable}}}}$ &
${}_{{{}\atop{\displaystyle\mbox{{\bf NB-by-MLE}$^{\,\rm a}$}}}}$  &
${}_{{{}\atop{\displaystyle\mbox{{\bf NB-by-MCMC}$^{\,\rm b}$}}}}$ &
\multicolumn{2}{|c|}{{\bf Restricted MSNB}$^{\,\rm c}$} &
\multicolumn{2}{|c|}{{\bf Full MSNB}$^{\,\rm d}$}
\\
\cline{4-7} & & & {\bf state }{\boldmath$s=0$} & {\bf state }{\boldmath$s=1$} &
{\bf state }{\boldmath$s=0$} & {\bf state }{\boldmath$s=1$}
\\
\hline
Intercept (constant term) &
$-21.3^{-18.7}_{-23.9}$ &
$-20.6^{-18.5}_{-22.7}$ &
$-20.9^{-18.7}_{-23.0}$ &
$-19.9^{-17.8}_{-22.1}$ &
$-20.7^{-18.7}_{-22.8}$ &
$-20.7^{-18.7}_{-22.8}$\\
\hline
Accident occurring on interstates I-70 or I-164 (dummy) &
$-.655^{-.562}_{-.748}$ &
$-.657^{-.565}_{-.750}$ &
$-.656^{-.564}_{-.748}$ &
$-.656^{-.564}_{-.748}$ &
$-.660^{-.568}_{-.752}$ &
$-.660^{-.568}_{-.752}$\\
\hline
Pavement quality index (PQI) average$^{\,\rm e}$ &
$-.0132^{-.00581}_{-.0205}$ &
$-.0189^{-.0134}_{-.0244}$ &
$-.0195^{-.0141}_{-.0248}$ &
$-.0195^{-.0141}_{-.0248}$ &
$-.0220^{-.0166}_{-.0273}$ &
$-.0125^{-.00700}_{-.0180}$\\
\hline
Road segment length (in miles) &
$.0512^{.0809}_{.0215}$ &
$.0546^{.0826}_{.0266}$ &
$.0538^{.0812}_{.0264}$ &
$.0538^{.0812}_{.0264}$ &
$.0395^{.0625}_{.0165}$ &
$.0395^{.0625}_{.0165}$\\
\hline
Logarithm of road segment length (in miles) &
$.909^{.974}_{.845}$ &
$.903^{.964}_{.842}$ &
$.900^{.961}_{.840}$ &
$.900^{.961}_{.840}$ &
$.913^{.973}_{.853}$ &
$.913^{.973}_{.853}$\\
\hline
Total number of ramps on the road viewing and opposite sides &
$-.0172^{-.00174}_{-.0327}$ &
$-.021^{-.00624}_{-.0358}$ &
$-.0187^{-.00423}_{-.0331}$ &
$-.0187^{-.00423}_{-.0331}$ &
-- &
$-.0264^{-.00656}_{-.0464}$\\
\hline
Number of ramps on the viewing side per lane per mile &
$.394^{.479}_{.309}$ &
$.400^{.479}_{.319}$ &
$.397^{.475}_{.317}$ &
$.397^{.475}_{.317}$ &
$.359^{.429}_{.289}$ &
$.359^{.429}_{.289}$\\
\hline
Median configuration is depressed (dummy) &
$.210^{.314}_{.106}$ &
$.214^{.318}_{.111}$ &
$.211^{.315}_{.108}$ &
$.211^{.315}_{.108}$ &
$.209^{.313}_{.107}$ &
$.209^{.313}_{.107}$\\
\hline
Median barrier presence (dummy) &
$-3.02^{-2.38}_{-3.67}$ &
$-2.99^{-2.40}_{-3.67}$ &
$-3.01^{-2.42}_{-3.69}$ &
$-3.01^{-2.42}_{-3.69}$ &
$-3.01^{-2.42}_{-3.69}$ &
$-3.01^{-2.42}_{-3.69}$\\
\hline
Interior shoulder presence (dummy) &
$-1.15^{-.486}_{-1.81}$ &
$-1.06^{.135}_{-2.26}$ &
$-1.02^{.148}_{-2.23}$ &
$-1.02^{.148}_{-2.23}$ &
$-1.16^{-.523}_{-1.87}$ &
$-1.16^{-.523}_{-1.87}$\\
\hline
Width of the interior shoulder is less that 5 feet (dummy) &
$.373^{.477}_{.270}$ &
$.384^{.491}_{.279}$ &
$.386^{.492}_{.281}$ &
$.386^{.492}_{.281}$ &
$.380^{.486}_{.275}$ &
$.380^{.486}_{.275}$\\
\hline
Interior rumble strips presence (dummy) &
$-.166^{-.0382}_{-.293}$ &
$-.142^{.857}_{-1.16}$ &
$-.163^{.836}_{-1.14}$ &
$-.163^{.836}_{-1.14}$ &
-- &
--\\
\hline
Width of the outside shoulder is less that 12 feet (dummy) &
$.281^{.380}_{.182}$ &
$.272^{.370}_{.174}$ &
$.268^{.366}_{.170}$ &
$.268^{.366}_{.170}$ &
$.267^{.365}_{.170}$ &
$.267^{.365}_{.170}$\\
\hline
Outside barrier is absent (dummy) &
$-.249^{-.139}_{-.358}$ &
$-.255^{-.142}_{-.366}$ &
$-.255^{-.142}_{-.366}$ &
$-.255^{-.142}_{-.366}$ &
$-.251^{-.140}_{-.362}$ &
$-.251^{-.140}_{-.362}$\\
\hline
Average annual daily traffic (AADT) &
$\displaystyle{-4.09^{-3.04\vphantom{O^o}}_{-5.15}}\atop\displaystyle{{}\times10^{-5}}$ &
$\displaystyle{-4.09^{-3.24\vphantom{O^o}}_{-4.95}}\atop\displaystyle{{}\times10^{-5}}$ &
$\displaystyle{-4.07^{-3.22\vphantom{O^o}}_{-4.94}}\atop\displaystyle{{}\times10^{-5}}$ &
$\displaystyle{-4.07^{-3.22\vphantom{O^o}}_{-4.94}}\atop\displaystyle{{}\times10^{-5}}$ &
$\displaystyle{-3.90^{-3.11\vphantom{O^o}}_{-4.72}}\atop\displaystyle{{}\times10^{-5}}$ &
$\displaystyle{-4.53^{-3.61\vphantom{O^o}}_{-5.48}}\atop\displaystyle{{}\times10^{-5}}$\\
\hline
Logarithm of average annual daily traffic  &
$2.08^{2.36}_{1.80}$ &
$2.06^{2.30}_{1.83}$ &
$2.07^{2.30}_{1.83}$ &
$2.07^{2.30}_{1.83}$ &
$2.07^{2.30}_{1.84}$ &
$2.07^{2.30}_{1.84}$\\
\hline
Posted speed limit (in mph) &
$.0154^{.0244}_{.00643}$ &
$.0150^{.0241}_{.00589}$ &
$.0161^{.0251}_{.00697}$ &
$.0161^{.0251}_{.00697}$ &
$.0161^{.0252}_{.00712}$ &
$.0161^{.0252}_{.00712}$\\
\hline
Number of bridges per mile &
$-.0213^{-.00187}_{-.0407}$ &
$-.0241^{-.00721}_{-.0419}$ &
$-.0233^{-.00648}_{-.0410}$ &
$-.0233^{-.00648}_{-.0410}$ &
-- &
$-.0607^{-.0232}_{-.102}$\\
\hline
Maximum of reciprocal values of horizontal curve radii (in $1/{\rm mile}$) &
$-.182^{-.122}_{-.242}$ &
$-.179^{-.118}_{-.241}$ &
$-.178^{-.117}_{-.239}$ &
$-.178^{-.117}_{-.239}$ &
$-.175^{-.114}_{-.237}$ &
$-.175^{-.114}_{-.237}$\\
\hline
Maximum of reciprocal values of vertical curve radii (in $1/{\rm mile}$) &
$.0191^{.0285}_{.00972}$ &
$.0177^{.027}_{.00843}$ &
$.0183^{.0275}_{.00917}$ &
$.0183^{.0275}_{.00917}$ &
$.0184^{.0274}_{.00925}$ &
$.0184^{.0274}_{.00925}$\\
\hline
Number of vertical curves per mile &
$-.0535^{-.0180}_{-.0889}$ &
$-.057^{-.0233}_{-.0924}$ &
$-.0586^{-.0249}_{-.0940}$ &
$-.0586^{-.0249}_{-.0940}$ &
$-.0565^{-.0231}_{-.0917}$ &
$-.0565^{-.0231}_{-.0917}$\\
\hline
Percentage of single unit trucks (daily average) &
$1.38^{1.88}_{.886}$ &
$1.25^{1.75}_{.758}$ &
$1.19^{1.68}_{.701}$ &
$1.19^{1.68}_{.701}$ &
$.726^{1.28}_{.171}$ &
$2.57^{3.39}_{1.77}$\\
\hline
\end{tabular}

\end{scriptsize}
\end{table}
\addtocounter{table}{-1}
\begin{table}[p]
\caption{(Continued)}
\begin{scriptsize}

\tabcolsep=0.45em
\renewcommand{\arraystretch}{1.45}
\begin{tabular}{|l|c|c|c|c|c|c|}
\hline
${}_{{{}\atop{\displaystyle\mbox{\bf Variable}}}}$ &
${}_{{{}\atop{\displaystyle\mbox{{\bf NB-by-MLE}$^{\,\rm a}$}}}}$  &
${}_{{{}\atop{\displaystyle\mbox{{\bf NB-by-MCMC}$^{\,\rm b}$}}}}$ &
\multicolumn{2}{|c|}{{\bf Restricted MSNB}$^{\,\rm c}$} &
\multicolumn{2}{|c|}{{\bf Full MSNB}$^{\,\rm d}$}
\\
\cline{4-7} & & & {\bf state }{\boldmath$s=0$} & {\bf state }{\boldmath$s=1$} &
{\bf state }{\boldmath$s=0$} & {\bf state }{\boldmath$s=1$}
\\
\hline
Winter season (dummy) &
$.148^{.226}_{.0698}$ &
$.148^{.226}_{.0689}$ &
$-.116^{.0563}_{-.261}$ &
$-.116^{.0563}_{-.261}$ &
$-.159^{-.0494}_{-.269}$ &
--\\
\hline
Spring season (dummy) &
$-.173^{-.0878}_{-.258}$ &
$-.173^{-.0899}_{-.257}$ &
$-.0932^{.0547}_{-.209}$ &
$-.0932^{.0547}_{-.209}$ &
-- &
--\\
\hline
Summer season (dummy) &
$-.179^{-.0921}_{-.266}$ &
$-.180^{-.0963}_{-.263}$ &
$-.0332^{.111}_{-.146}$ &
$-.0332^{.111}_{-.146}$ &
-- &
$-.549^{-.293}_{-.883}$\\
\hline
Over-dispersion parameter $\alpha$ in NB models &
$.957^{1.07}_{.845}$ &
$.968^{1.09}_{.849}$ &
$.537^{.677}_{.392}$ &
$1.24^{1.51}_{.986}$ &
$.443^{.595}_{.300}$ &
$1.16^{1.39}_{.945}$\\
\hline
\hline
Mean accident rate ($\lambda_{t,n}$ for NB),
averaged over all values of ${\bf X}_{t,n}$
& -- &
$.0663$ &
$.0558$ &
$.1440$ &
$.0533$ &
$.1130$\\
\hline
Standard deviation of accident rate
($\sqrt{\lambda_{t,n}(1+\alpha\lambda_{t,n})}\,$ for NB), & & & & & &\\
averaged over all values of explanatory variables ${\bf X}_{t,n}$ & -- &
$.2050$ &
$.1810$ &
$.3350$ &
$.1760$ &
$.2820$\\
\hline
Markov transition probability of jump $0\to1$ ($p_{0\rightarrow1}$) &
-- & -- &
\multicolumn{2}{|c|}{$.0933^{.147}_{.0531}$} &
\multicolumn{2}{|c|}{$.158^{.225}_{.100}$}\\
\hline
Markov transition probability of jump $1\to0$ ($p_{1\rightarrow0}$) &
-- & -- &
\multicolumn{2}{|c|}{$.651^{.820}_{.463}$} &
\multicolumn{2}{|c|}{$.627^{.773}_{.474}$}\\
\hline
Unconditional probabilities of states 0 and 1 (${\bar p}_0$ and ${\bar p}_1$)
& -- & -- &
\multicolumn{2}{|c|}{$.873^{.929}_{.797}$\quad and\quad $.127^{.203}_{.0713}$} &
\multicolumn{2}{|c|}{$.798^{.868}_{.718}$\quad and\quad $.202^{.282}_{.132}$}\\
\hline
Total number of free model parameters ($\beta$-s and $\alpha$-s) &
$26$ & $26$ &
\multicolumn{2}{|c|}{$28$} &
\multicolumn{2}{|c|}{$28$}\\
\hline
Posterior average of the log-likelihood (LL) & -- &
$-16097.2^{-16091.3}_{-16105.0}$ &
\multicolumn{2}{|c|}{$-15821.8^{-15807.9}_{-15835.2}$} &
\multicolumn{2}{|c|}{$-15778.0^{-15672.9}_{-15794.9}$}\\
\hline
Max$(LL)$:\quad estimated max.~value of log-likelihood (LL) for MLE; & & &
\multicolumn{2}{|c|}{} & \multicolumn{2}{|c|}{} \\
max. observed LL during MCMC simulations for Bayesian estim. &
$-16081.2\,{\rm(MLE)}\!$ &
$-16086.3\,{\rm(observ.)}\!$ &
\multicolumn{2}{|c|}{$-15786.6\,$(observed)} &
\multicolumn{2}{|c|}{$-15744.8\,$(observed)}\\
\hline
Logarithm of marginal likelihood of data ($\ln[f({\bf Y}|{\cal M})]$) &
-- &
$-16108.6^{-16105.7}_{-16110.7}$ &
\multicolumn{2}{|c|}{$-15850.2^{-15840.1}_{-15849.5}$} &
\multicolumn{2}{|c|}{$-15809.4^{-15801.7}_{-15811.9}$}\\
\hline
Goodness-of-fit p-value &
-- & $0.701$ &
\multicolumn{2}{|c|}{$0.729$} &
\multicolumn{2}{|c|}{$0.647$}\\
\hline
Maximum of the potential scale reduction factors (PSRF)$^{\,\rm f}$ & -- &
$1.00874$ &
\multicolumn{2}{|c|}{$1.00754$} &
\multicolumn{2}{|c|}{$1.00939$}\\
\hline
Multivariate potential scale reduction factor (MPSRF)$^{\,\rm f}$ &
-- &
$1.00928$ &
\multicolumn{2}{|c|}{$1.00925$} &
\multicolumn{2}{|c|}{$1.01002$}\\
\hline
\multicolumn{7}{l}{$^{\rm a}$ Standard (conventional) negative binomial
estimated by maximum likelihood estimation (MLE).}\\
\multicolumn{7}{l}{$^{\rm b}$ Standard negative binomial
estimated by Markov Chain Monte Carlo (MCMC) simulations.}\\
\multicolumn{7}{l}{$^{\rm c}$ Restricted two-state Markov switching negative
binomial (MSNB) model with only the intercept and over-dispersion parameters
allowed to vary between states.}\\
\multicolumn{7}{l}{$^{\rm d}$ Full two-state Markov switching negative
binomial (MSNB) model with all parameters allowed to vary between states.}\\
\multicolumn{7}{l}{$^{\rm e}$ The pavement quality index (PQI) is a composite
measure of overall pavement quality evaluated on a 0 to 100 scale.}\\
\multicolumn{7}{l}{$^{\rm f}$ PSRF/MPSRF are calculated separately/jointly
for all continuous model parameters. PSRF and MPSRF are close to 1 for
converged MCMC chains.}
\end{tabular}

\end{scriptsize}
\end{table}
\begin{table}[p]
\caption{Summary statistics of roadway segment characteristic variables}
\label{T_2}
\begin{scriptsize}

\tabcolsep=0.45em
\renewcommand{\arraystretch}{1.45}
\begin{tabular}{|l|c|c|c|c|c|}
\hline
{\bf Variable} &
{\bf Mean} &
{\bf Standard deviation} &
{\bf Minimum} &
{\bf Median} &
{\bf Maximum}
\\
\hline
Accident occurring on interstates I-70 or I-164 (dummy) &
$.155$ &
$.363$ &
$0$ &
$0$ &
$1.00$
\\
\hline
Pavement quality index (PQI) average$^{\,\rm a}$ &
$88.6$ &
$5.96$ &
$69.0$ &
$90.3$ &
$98.5$
\\
\hline
Road segment length (in miles) &
$.886$ &
$1.48$ &
$.00900$ &
$.356$ &
$11.5$
\\
\hline
Logarithm of road segment length (in miles) &
$-.901$ &
$1.22$ &
$-4.71$ &
$-1.03$ &
$2.44$
\\
\hline
Total number of ramps on the road viewing and opposite sides &
$.725$ &
$1.79$ &
$0$ &
$0$ &
$16$
\\
\hline
Number of ramps on the viewing side per lane per mile &
$.138$ &
$.408$ &
$0$ &
$0$ &
$3.27$
\\
\hline
Median configuration is depressed (dummy) &
$.630$ &
$.484$ &
$0$ &
$1.00$ &
$1.00$
\\
\hline
Median barrier presence (dummy) &
$.161$ &
$.368$ &
$0$ &
$0$ &
$1$
\\
\hline
Interior shoulder presence (dummy) &
$.928$ &
$.258$ &
$0$ &
$1$ &
$1$
\\
\hline
Width of the interior shoulder is less that 5 feet (dummy) &
$.696$ &
$.461$ &
$0$ &
$1.00$ &
$1.00$
\\
\hline
Interior rumble strips presence (dummy) &
$.722$ &
$.448$ &
$0$ &
$1.00$ &
$1.00$
\\
\hline
Width of the outside shoulder is less that 12 feet (dummy) &
$.752$ &
$.432$ &
$0$ &
$1.00$ &
$1.00$
\\
\hline
Outside barrier absence (dummy) &
$.830$ &
$.376$ &
$0$ &
$1.00$ &
$1.00$
\\
\hline
Average annual daily traffic (AADT) &
$3.03\times10^4$ &
$2.89\times10^4$ &
$.944\times10^4$ &
$1.65\times10^4$ &
$14.3\times10^4$
\\
\hline
Logarithm of average annual daily traffic  &
$10.0$ &
$.623$ &
$9.15$ &
$9.71$ &
$11.9$
\\
\hline
Posted speed limit (in mph) &
$63.1$ &
$3.89$ &
$50.0$ &
$65.0$ &
$65.0$
\\
\hline
Number of bridges per mile &
$1.76$ &
$8.14$ &
$0$ &
$0$ &
$124$
\\
\hline
Maximum of reciprocal values of horizontal curve radii (in $1/{\rm mile}$) &
$.650$ &
$.632$ &
$0$ &
$.589$ &
$2.26$
\\
\hline
Maximum of reciprocal values of vertical curve radii (in $1/{\rm mile}$) &
$2.38$ &
$3.59$ &
$0$ &
$0$ &
$14.9$
\\
\hline
Number of vertical curves per mile &
$1.50$ &
$4.03$ &
$0$ &
$0$ &
$50.0$
\\
\hline
Percentage of single unit trucks (daily average) &
$.0859$ &
$.0678$ &
$.00975$ &
$.0683$ &
$.322$
\\
\hline
Winter season (dummy) &
$.242$ &
$.428$ &
$0$ &
$0$ &
$1.00$
\\
\hline
Spring season (dummy) &
$.254$ &
$.435$ &
$0$ &
$0$ &
$1.00$
\\
\hline
Summer season (dummy) &
$.254$ &
$.435$ &
$0$ &
$0$ &
$1.00$
\\
\hline
\multicolumn{6}{l}{$^{\rm a}$ The pavement quality index (PQI) is a composite
measure of overall pavement quality evaluated on a 0 to 100 scale.}
\end{tabular}

\end{scriptsize}
\end{table}
\end{landscape}

Focusing on the full MSNB model, which is statistically superior,
its estimation results show that the less frequent state $s_t=1$
is about four times as rare as the more frequent state $s_t=0$
(refer to the estimated values of the unconditional probabilities
${\bar p}_0$ and ${\bar p}_1$ of states $0$ and $1$, which are
given in the ``Full MSNB'' columns in Table~\ref{T_1}).

Also, the findings show that the less frequent state $s_t=1$
is considerably less safe than the more frequent state $s_t=0$.
This result follows from the values of the mean weekly accident
rate $\lambda_{t,n}$ [given by Eq.~(\ref{EQ_LAMBDA}) with model
parameters $\beta$-s set to their posterior means in the
two states], averaged over all values of the
explanatory variables ${\bf X}_{t,n}$ observed in the data
sample (see ``mean accident rate'' in Table~\ref{T_1}). For the
full MSNB model, on average, state $s_t=1$ has about two times
more accidents per week than state $s_t=0$ has.\footnote{Note that
accident frequency rates can easily be converted from one time period
to another (for example, weekly rates can be converted to annual
rates). Because accident events are independent, the conversion is done
by a summation of moment-generating (or characteristic) functions.
The sum of Poisson variates is Poisson. The sum of NB variates is
also NB if all explanatory variables do not depend on time
(${\bf X}_{t,n}={\bf X}_n$).}
Therefore, it is not a surprise, that in Figure~\ref{FIGURE_2} the
weekly number of accidents (shown on the bottom plot) is larger when the
posterior probability $P(s_t=1|{\bf Y})$ of the state $s_t=1$ (shown
on the top plot) is higher.
Note that the long-term unconditional mean of the accident rates is
equal to the average of the mean accident rate over the two states,
this average is calculated by using the stationary probabilities
$\bar p_{0}$ and $\bar p_{1}$ given by Eq.~(\ref{EQ_P_BAR}) (see
the ``unconditional probabilities of states 0 and 1'' in
Table~\ref{T_1}).

It is also noteworthy that the number of accidents is more volatile
in the less frequent and less-safe state ($s_t=1$).
This is reflected
in the fact that the standard deviation of the accident rate
(${\rm std}_{t,n}=\sqrt{\lambda_{t,n}(1+\alpha\lambda_{t,n})}$
for NB distribution), averaged over all
values of explanatory variables ${\bf X}_{t,n}$, is higher in
state $s_t=1$ than in state $s_t=0$ (refer to Table~\ref{T_1}).
Moreover, for the full MSNB model the over-dispersion parameter
$\alpha$ is higher in state $s_t=1$
($\alpha=0.443$ in state $s_t=0$ and $\alpha=1.16$ in state
$s_t=1$). Because state $s_t=1$ is relatively rare, this suggests
that over-dispersed volatility of accident frequencies, which is
often observed in empirical data, could be in part due to the
latent switching between the states, and in part due to high
accident volatility in the less frequent and less safe state
$s_t=1$.

To study the effect of weather (which is usually
unobserved heterogeneity in most data bases) on states,
Table~\ref{T_3} gives time-correlation coefficients
between posterior probabilities $P(s_t=1|{\bf Y})$ for the full
MSNB model and weather-condition variables. These correlations
were found by using daily and hourly historical weather data in
Indiana, available at the Indiana State Climate Office at Purdue
University (www.agry.purdue.edu/climate). For these correlations,
the precipitation and snowfall amounts are daily amounts in inches
averaged over the week and across several weather observation
stations that are located close to the roadway
segments.\footnote{Snowfall and precipitation amounts
are weakly related with each other because snow density $(g/cm^3)$
can vary by more than a factor of ten.} The temperature variable
is the mean daily air temperature $(^oF)$ averaged over the week
and across the weather stations.
The effect of fog/frost is captured by a dummy variable that is
equal to one if and only if the difference between air and dewpoint
temperatures does not exceed $5^oF$ (in this case frost can form if
the dewpoint is below the freezing point $32^oF$, and fog can form
otherwise). The fog/frost dummies are calculated for every hour and
are averaged over the week and across the weather stations.
Finally, visibility distance variable is the harmonic mean of hourly
visibility distances, which are measured in miles every hour and are
averaged over the week and across the weather stations.\footnote{The
harmonic mean $\bar d$ of distances $d_n$ is calculated as
$\bar d^{-1}=(1/N)\sum^N_{n=1}d^{-1}_n$, assuming $d_n=0.25$ miles if
$d_n\leq0.25$ miles.}

\begin{table}[t]
\caption{Correlations of the posterior probabilities $P(s_t=1|{\bf Y})$
with weather-condition variables for the full MSNB model}
\label{T_3}
\begin{footnotesize}

\begin{tabular}{|l|c|c|c|}
\hline
& {\bf All year} & {\bf Winter} & {\bf Summer}\\
& & (November--March) & (May--September)\\
\hline
Precipitation (inch) & $0.031$ & -- & $0.144$\\
\hline
Temperature ($^oF$) & $-0.518$ & $-0.591$ & $0.201$\\
\hline
Snowfall (inch) & $0.602$ & $0.577$ & --\\
\cline{2-4}
$\quad\quad\quad>0.2$ (dummy) & $0.651$ & $0.638$ & --\\
\hline
Fog / Frost (dummy) & $0.223$ & (frost) $0.539$ & (fog) $0.051$\\
\hline
Visibility distance (mile) & $-0.221$ & $-0.232$ & $-0.126$\\
\hline
\end{tabular}

\end{footnotesize}
\end{table}

Table~\ref{T_3} shows that the less frequent and less safe state $s_t=1$
is positively correlated with extreme temperatures (low during winter
and high during summer), rain precipitations and snowfalls, fogs
and frosts, low visibility distances. It is reasonable to expect
that during bad weather, roads can become significantly less safe,
resulting in a change of the state of roadway safety.
As a useful test of the switching between the two states, all weather
variables, listed in Table~\ref{T_3}, were added into our full
MSNB model. However, when doing this, the two states did not
disappear and the posterior probabilities $P(s_t=1|{\bf Y})$ did not
changed substantially (the correlation between the new and the old
probabilities was around $90\%$).

In addition to the MSNB models,
we estimated two-state Markov switching Poisson (MSP) models, which
have the Poisson likelihood function instead of the NB likelihood
function in Eq.~(\ref{EQ_L_1}). Our findings for the MSP models are
very similar to those for the MSNB models \citep[][]{M_08}. Also,
because the time series in Figure~\ref{FIGURE_2} seems to exhibit
a seasonal pattern (roads appear to be less safe and $P(s_t=1|{\bf Y})$
appears to be higher during winters), we estimated MSNB and MSP
models in which the transition probabilities $p_{0\rightarrow1}$
and $p_{1\rightarrow0}$ are not constant (allowing each of them to
assume two different values: one during winters and the other during
all remaining seasons).
However, these models did not perform as well
as the MSNB and MSP models with constant transition probabilities
[as judged by the Bayes factors, see
Eq.~(\ref{EQ_BAYES_FACTOR})].\footnote{We have only five winter
periods in our five-year data. MSNB and MSP with seasonally
changing transition probabilities could perform better for an
accident data that covers a longer time interval.}

\section{Summary and conclusions}
\label{CONCLUD}

The empirical finding that two states exist and that these states
are correlated with weather conditions has important implications.
The findings suggest that multiple states of roadway safety can exist
due to slow and/or inadequate adjustment by drivers (and possibly by
roadway maintenance services) to adverse conditions and other
unpredictable, unidentified, and/or unobservable variables that
influence roadway safety. All these variables are likely
to interact and change over time, resulting in transitions from one
state to the next.

As discussed earlier, the empirical findings show that the less
frequent state is significantly less safe than the other, more
frequent state. The full MSNB model results show that explanatory
variables ${\bf X}_{t,n}$, other than
the intercept, exert different influences on roadway safety in
different states as indicated by the fact that some of the parameter
estimates for the two states of the full MSNB model are significantly
different.\footnote{Table~\ref{T_1} shows that parameter estimates for
pavement quality index, total number of ramps on the road viewing
and opposite sides, average annual daily traffic, number of bridges
per mile, and percentage of single unit trucks are all significantly
different between the two states.}
Thus, the states not only differ by average accident frequencies, but
also differ in the magnitude and/or direction of the effects that
various variables exert on accident frequencies. This again
underscores the importance of the two-state approach.

The Markov switching models presented in this study are similar to
zero-inflated count data models (which have been previously applied
in accident frequency research) in the sense that they are also
two-state models \citep[see][]{SMM_97,LWI_07}. However, in contrast
to zero-inflated models, the models presented herein allow for
switching between the two states over time. Also, in this study, a
``safe'' state is not assumed and accident frequencies can be
nonzero in both states.

In terms of future work on Markov switching models for roadway safety,
additional empirical studies (for other accident data samples), and
multi-state models (with more than two states of roadway safety) are
two areas that would further demonstrate the potential of the approach.

\section*{Acknowledgments}
\label{ACKN}

We thank Dominique Lord for his interest in this study and for useful discussions.


\appendix
\section{$\!\!\!\!\!\!\!$ppendix: MCMC simulation algorithm}
\label{APPENDIX}

For brevity, in this appendix we omit model notation ${\cal M}$ in all
equations. For example, here we write the posterior distribution,
given by Eq.~(\ref{EQ_POSTERIOR}), as $f({\bf\Theta}|{\bf Y})$.

We choose the prior distribution $\pi({\bf\Theta})$ of parameter
vector ${\bf\Theta}$, given by Eq.~(\ref{EQ_THETA}), as
\begin{eqnarray}
&& \!\!\! \pi({\bf\Theta})=f({\bf S}|p_{0\rightarrow1},p_{1\rightarrow0})
\pi(p_{0\rightarrow1},p_{1\rightarrow0})
\prod\limits_{s=0}^1\left[\pi(\alpha_{(s)})\prod\limits_k\pi(\beta_{(s),k})\right],
\label{EQA_PRIOR}\\
&& \!\!\! \pi(p_{0\rightarrow1},p_{1\rightarrow0})\propto
\pi(p_{0\rightarrow1})\pi(p_{1\rightarrow0})
I(p_{0\rightarrow1}\le p_{1\rightarrow0}),
\label{EQA_PRIOR_P}\\
&& \!\!\! f({\bf S}|p_{0\rightarrow1},p_{1\rightarrow0})=P(s_1)\prod\limits_{t=2}^T
P(s_t|s_{t-1})\propto\prod\limits_{t=2}^T P(s_t|s_{t-1})={}
\nonumber\\
&& \!\!\! \qquad\qquad\quad =
(p_{0\rightarrow1})^{n_{0\rightarrow1}}(1-p_{0\rightarrow1})^{n_{0\rightarrow0}}
(p_{1\rightarrow0})^{n_{1\rightarrow0}}(1-p_{1\rightarrow0})^{n_{1\rightarrow1}}
\label{EQA_PRIOR_S}
\end{eqnarray}
Here $\beta_{(s),k}$ is the $k^{\rm th}$ component of vector
$\betabf_{(s)}$. The priors of $\alpha_{(s)}$ and $\beta_{(s),k}$
are chosen to be normal distributions,
$\pi(\alpha_{(s)})={\cal N}(\mu_\alpha,\Sigma_\alpha)$ and
$\pi(\beta_{(s),k})={\cal N}(\mu_k,\Sigma_k)$.
The joint prior of $p_{0\rightarrow1}$ and $p_{1\rightarrow0}$
is given by Eq.~(\ref{EQA_PRIOR_P}), where
$\pi(p_{0\rightarrow1})={\cal B}eta(\upsilon_0,\nu_0)$ and
$\pi(p_{1\rightarrow0})={\cal B}eta(\upsilon_1,\nu_1)$ are beta
distributions, and function $I(p_{0\rightarrow1}\le p_{1\rightarrow0})$
is equal to unity if the restriction given by Eq.~(\ref{EQ_P_RESTRICT})
is satisfied and to zero otherwise. We disregard distribution
$P(s_1)$ in Eq.~(\ref{EQA_PRIOR_S}) because its contribution
is negligible when $T$ is large (alternatively, we can assume
$P(s_1=0)=P(s_1=1)=1/2$). Number $n_{i\rightarrow j}$ is the
total number of state transitions $i\rightarrow j$ (where
$i,j=0,1$). Parameters that enter the prior distributions are called
hyper-parameters. For these, the means $\mu_\alpha$ and $\mu_k$
are equal to the maximum likelihood estimation (MLE) values of
$\alpha$ and $\beta_k$ for the corresponding standard single-state
model (NB model in this study). The variances $\Sigma_\alpha$ and
$\Sigma_k$ are ten times larger than the maximum between the MLE
values of $\alpha$ and $\beta_k$ squared and the MLE variances of
$\alpha$ and $\beta_k$ for the standard model. We choose
$\upsilon_0=\nu_0=\upsilon_1=\nu_1=1$ (in this case the beta
distributions become the uniform distribution between zero and one).

To obtain draws from a posterior distribution, we use the hybrid Gibbs
sampler, which is an MCMC simulation algorithm that involves both Gibbs
and Metropolis-Hasting sampling \citep[][]{MT_94,T_02,SAS_06}. Assume
that ${\bf\Theta}$ is composed of $K$ components:
${\bf\Theta}=[\thetabf_1',\thetabf_2',\ldots,\thetabf_K']'$ , where
$\thetabf_k$ can be scalars or vectors, $k=1,2,\ldots,K$. Then, the
hybrid Gibbs sampler works as follows:
\begin{enumerate}
    \item Choose an arbitrary initial value of the parameter vector,
      ${\bf\Theta}={\bf\Theta}^{(0)}$ , such that
      $f({\bf Y},{\bf\Theta}^{(0)})>0$.
    \item For each $g=1,2,3,\ldots\;$, parameter vector
      ${\bf\Theta}^{(g)}$ is generated component-by-component from
      ${\bf\Theta}^{(g-1)}$ by the following procedure:
    \begin{enumerate}
      \item First, draw $\thetabf^{(g)}_1$ from the conditional
        posterior probability distribution
        $f(\thetabf^{(g)}_1|{\bf Y},\thetabf^{(g-1)}_2,\ldots,\thetabf^{(g-1)}_K)$.
        If this distribution is exactly known in a closed analytical
        form, then we draw $\thetabf^{(g)}_1$ directly from it. This
        is Gibbs sampling. If the conditional posterior distribution
        is known up to an unknown normalization constant, then we
        draw $\thetabf^{(g)}_1$ by using the Metropolis-Hasting
        ($\mbox{M-H}$) algorithm described below. This is $\mbox{M-H}$
        sampling.
      \item Second, for all $k=2,3,\ldots,K-1$, draw $\thetabf^{(g)}_k$
        from the conditional posterior distribution
        $f(\thetabf^{(g)}_k|{\bf Y}, \thetabf^{(g)}_1,\ldots,
        \thetabf^{(g)}_{k-1},\thetabf^{(g-1)}_{k+1},\ldots,\thetabf^{(g-1)}_K)$
        by using either Gibbs sampling (if the distribution is known
        exactly) or $\mbox{M-H}$ sampling (if the distribution is
        known up to a constant).
      \item Finally, draw $\thetabf^{(g)}_K$ from conditional
        posterior probability distribution
        $f(\thetabf^{(g)}_K|{\bf Y},\thetabf^{(g)}_1,\ldots,\thetabf^{(g)}_{K-1})$
        by using either Gibbs or $\mbox{M-H}$ sampling.
    \end{enumerate}
    \item The resulting Markov chain $\{{\bf\Theta}^{(g)}\}$ converges
      to the true posterior distribution $f({\bf\Theta}|{\bf Y})$ as
      $g\rightarrow\infty$.
\end{enumerate}
Note that all conditional posterior distributions are proportional
to the joint distribution
$f({\bf Y},{\bf\Theta})=f({\bf Y}|{\bf\Theta})\pi({\bf\Theta})$,
where the likelihood $f({\bf Y}|{\bf\Theta})$ is given by
Eq.~(\ref{EQ_L}) and the prior $\pi({\bf\Theta})$ is given by
Eq.~(\ref{EQA_PRIOR}).

By using the hybrid Gibbs sampler algorithm described above, we obtain
a Markov chain $\{{\bf\Theta}^{(g)}\}$, where
$g=1,2,\ldots,G_{bi},G_{bi}+1,\ldots,G$. We discard the first $G_{bi}$
``burn-in'' draws because they can depend on the initial choice
${\bf\Theta}^{(0)}$. Of the remaining $G-G_{bi}$ draws, we typically
store every third or every tenth draw in the computer memory. We use
these draws for Bayesian inference. Our typical choice is
$G_{bi}=3\times10^5$ and $G=3\times10^6$. In our study, one MCMC
simulation run takes few days on a single computer CPU. We usually
consider eight choices of the initial parameter vector
${\bf\Theta}^{(0)}$. Thus, we obtain eight Markov chains of
${\bf\Theta}$, and use them for the Brooks-Gelman-Rubin diagnostic of
convergence of our MCMC simulations \citep[][]{BG_98}. We also check
convergence by monitoring the likelihood
$f({\bf Y}|{\bf\Theta}^{(g)})$ and the joint distribution
$f({\bf Y},{\bf\Theta}^{(g)})$.

The Metropolis-Hasting ($\mbox{M-H}$) algorithm is used to sample
from conditional posterior distributions known up to their
normalization constants.\footnote{In general, the M-H algorithm allows
to make draws from any probability distribution known up to a constant.
The algorithm converges as the number of draws goes to infinity.}
For brevity, we use notation
$f_g(\thetabf_k|{\bf Y},{\bf\Theta}\backslash\thetabf_k)\equiv
f(\thetabf_k|{\bf Y},\thetabf^{(g)}_1,\ldots,\thetabf^{(g)}_{k-1},
\thetabf^{(g-1)}_{k+1},\ldots,\thetabf^{(g-1)}_K)$, where
${\bf\Theta}\backslash\thetabf_k$ means all components of
${\bf\Theta}$ except $\thetabf_k$. The $\mbox{M-H}$ algorithm is as
follows:
\begin{itemize}
    \item Choose a jumping probability distribution
      $J(\hat\thetabf_k|\thetabf_k)$ of $\hat\thetabf_k$. It must stay
      the same for all draws $g=G_{bi}+1,\ldots,G$, and we discuss its
      choice below.
    \item Draw a candidate $\hat\thetabf_k$ from
      $J(\hat\thetabf_k|\thetabf_k^{(g-1)})$.
    \item Calculate ratio
        \begin{eqnarray}
        \hat p=\frac{f_g(\hat\thetabf_k|{\bf Y},{\bf\Theta}\backslash\thetabf_k)}
        {f_g(\thetabf_k^{(g-1)}|{\bf Y},{\bf\Theta}\backslash\thetabf_k)}\times
        \frac{J(\thetabf_k^{(g-1)}|\hat\thetabf_k)}
        {J(\hat\thetabf_k|\thetabf_k^{(g-1)})}.
        \label{EQA_MH_RATIO}
        \end{eqnarray}
    \item Set
        \begin{eqnarray}
        \thetabf_k^{(g)}=\left\{
            \begin{array}{ll}
            \hat\thetabf_k     & \mbox{with probability $\min(\hat p,1)$},\\
            \thetabf_k^{(g-1)} & \mbox{otherwise}.
            \end{array}
        \right.
        \label{EQA_MH_DRAW}
        \end{eqnarray}
\end{itemize}
Note that the unknown normalization constant of $f_g(\ldots)$ cancels
out in Eq.~(\ref{EQA_MH_RATIO}).

In this study ${\bf\Theta}$ is given by Eq.~(\ref{EQ_THETA}), and the
hybrid Gibbs sampler generates draws ${\bf\Theta}^{(g)}$ from
${\bf\Theta}^{(g-1)}$ as follows (for brevity, below we drop $g$
indexing):
\begin{enumerate}
    \item[(a)] We draw vector $\betabf_{(0)}$ component-by-component
      by using the $\mbox{M-H}$ algorithm. For each component
      $\beta_{(0),k}$ of $\betabf_{(0)}$ we use a normal jumping
      distribution
      $J(\hat\beta_{(0),k}|\beta_{(0),k})={\cal N}(\beta_{(0),k},\sigma^2_{(0),k})$.
      Variances $\sigma^2_{(0),k}$ are adjusted during the burn-in
      sampling ($g=1,2,\ldots,G_{bi}$) to have approximately $30\%$
      acceptance rate in Eq.~(\ref{EQA_MH_DRAW}).\footnote{We also
      tried Cauchy jumping distributions and obtained similar results.}
      The conditional posterior distribution of $\beta_{(0),k}$ is
      \begin{eqnarray}
      f(\beta_{(0),k}|{\bf Y},{\bf\Theta}\backslash\beta_{(0),k})\propto
      f({\bf Y},{\bf\Theta})=f({\bf Y}|{\bf\Theta})\pi({\bf\Theta})
      \propto f({\bf Y}|{\bf\Theta})\pi(\beta_{(0),k}).
      \nonumber
      \end{eqnarray}
    \item[(b)] We draw $\alpha_{(0)}$ first, all components of
      $\betabf_{(1)}$ second, and $\alpha_{(1)}$ third, from
      their conditional posterior distributions by using the
      $\mbox{M-H}$ algorithm in a way very similar to the drawing
      the components of $\betabf_{(0)}$. In all cases, we use normal
      jumping distributions with variances chosen to have
      $\approx30\%$ acceptance rates.
    \item[(c)] By using Gibbs sampling, we draw, first,
      $p_{0\rightarrow1}$ and, second, $p_{1\rightarrow0}$ from
      their conditional posterior distributions, which are
      truncated beta distributions,
      \begin{eqnarray}
      \begin{array}{l}
      f(p_{0\rightarrow1}|{\bf Y},{\bf\Theta}\backslash p_{0\rightarrow1})
      \propto f({\bf Y},{\bf\Theta})\propto
      f({\bf S}|p_{0\rightarrow1},p_{1\rightarrow0})
      \pi(p_{0\rightarrow1},p_{1\rightarrow0})\propto{}
      \\
      \qquad\qquad\quad
      {}\propto
      {\cal B}eta(\upsilon_0+n_{0\rightarrow1},\nu_0+n_{0\rightarrow0})
      I(p_{0\rightarrow1}\le p_{1\rightarrow0}),
      \\
      f(p_{1\rightarrow0}|{\bf Y},{\bf\Theta}\backslash p_{1\rightarrow0})
      \propto{\cal B}eta(\upsilon_1+n_{1\rightarrow0},\nu_1+n_{1\rightarrow1})
      I(p_{0\rightarrow1}\le p_{1\rightarrow0}).
      \end{array}
      \label{EQA_P_DRAW}
      \end{eqnarray}
    \item[(d)] Finally, we draw components of
      ${\bf S}=[s_1,s_2,\ldots,s_T]'$ by Gibbs sampling. Neighboring
      components of ${\bf S}$ can be strongly (anti-)correlated.
      Therefore, to speed up MCMC convergence in this case, we draw
      subsections ${\bf S}_{t,\tau}=[s_t,s_{t+1},\ldots,s_{t+\tau-1}]'$
      of ${\bf S}$ at a time. The conditional posterior distribution
      of ${\bf S}_{t,\tau}$ is
      \begin{eqnarray}
      f({\bf S}_{t,\tau}|{\bf Y},{\bf\Theta}\backslash{\bf S}_{t,\tau})\propto
      f({\bf Y},{\bf\Theta})\propto f({\bf Y}|{\bf\Theta})
      f({\bf S}|p_{0\rightarrow1},p_{1\rightarrow0}).
      \label{EQA_S_DRAWS}
      \end{eqnarray}
      Vector ${\bf S}_{t,\tau}$ has length $\tau$ and can assume
      $2^\tau$ possible values. By choosing $\tau$ small enough,
      we can compute the right-hand-side of Eq.~(\ref{EQA_S_DRAWS})
      for each of these values and find the normalization constant of
      $f({\bf S}_{t,\tau}|{\bf Y},{\bf\Theta}\backslash{\bf S}_{t,\tau})$.
      This allows us to make Gibbs sampling of ${\bf S}_{t,\tau}$.
      Our typical choice of $\tau$ is from 10 to 14. We draw all
      subsections ${\bf S}_{t,\tau}$ one after another.
\end{enumerate}

Additional details on MCMC simulation algorithms and their implementation
in the context of accident modeling can be found in \citet[][]{M_08}.

\end{document}